\documentclass{emulateapj}

\slugcomment{Accepted to ApJ on 16 March 2012}
\begin{document}

\title{Possible Detection of an Emission Cyclotron Resonance Scattering Feature from the Accretion-powered Pulsar 4U 1626$-$67}

\author{W.~B.~Iwakiri\altaffilmark{1},
Y.~Terada\altaffilmark{1},
T.~Mihara\altaffilmark{2},
L.Angelini\altaffilmark{3},
M.~S.~Tashiro\altaffilmark{1},
T.~Enoto\altaffilmark{4},
S.~Yamada\altaffilmark{3},
K.~Makishima\altaffilmark{5},
M.~Nakajima\altaffilmark{6},
A.~Yoshida\altaffilmark{7},
}

\altaffiltext{1}{Graduate School of Science and Engineering, Saitama University, 255 Shimo-Okubo, Sakura, Saitama 338-8570, Japan}
\altaffiltext{2}{Institute of Physical and Chemical Research (RIKEN), 2-1 Hirosawa, Wako, Saitama 351-0198, Japan}
\altaffiltext{3}{Laboratory for High Energy Astrophysics, NASA Goddard Space Flight Center, Code 660, Greenbelt, MD 20771}
\altaffiltext{4}{Kavli Institute for Particle Astrophysics and Cosmology,
Department of Physics and SLAC National Accelerator Laboratory,
Stanford University, Stanford, CA 94305, USA}
\altaffiltext{5}{Department of Physics, The University of Tokyo, 7-3-1 Hongo, Bunkyo, Tokyo 113-0033, Japan}
\altaffiltext{6}{School of Dentistry at Matsudo, Nihon University, 2-870-1 Sakaecho-Nishi, Matsudo, Chiba 271-8587, Japan}
\altaffiltext{7}{Department of Physics and Mathematics, Aoyama Gakuin University, 5-10-1 Fuchinobe, Sagamihara 229-8558, Japan}

\begin{abstract}
We present analysis of 4U 1626$-$67, a 7.7~s pulsar in a low-mass
X-ray binary system, observed with the hard X-ray detector of the
Japanese X-ray satellite {\it{Suzaku}} in March 2006 for a net
exposure of $\sim 88$ ks. The source was detected at an average 10--60
keV flux of $\sim 4 \times 10^{-10}$ erg cm$^{-2}$ s$^{-1}$. The
phase-averaged spectrum is reproduced well by combining a negative and
positive power-law times exponential cutoff (NPEX) model modified at
$\sim$ 37 keV by a cyclotron resonance scattering feature (CRSF). The
phase-resolved analysis shows that the spectra at the bright phases
are well fit by the NPEX with CRSF model. On the other hand, the
spectrum in the dim phase lacks the NPEX high-energy cutoff component,
and the CRSF can be reproduced by either an emission or an absorption
profile. When fitting the dim phase spectrum with the NPEX plus
Gaussian model, we find that the feature is better described in terms
of an emission rather than an absorption profile. The statistical
significance of this result, evaluated by means of an F-test, is
between $2.91 \times 10^{-3}$ and $1.53 \times 10^{-5}$, taking into
account the systematic errors in the background evaluation of
HXD-PIN. We find that, the emission profile is more feasible than the
absorption one for comparing the physical parameters in other
phases. Therefore, we have possibly detected an emission line at the
cyclotron resonance energy in the dim phase.

\end{abstract}


\section{Introduction}

Accreting binary pulsars are an ideal laboratory for studying the
fundamental physics of radiative transfer of X-ray photons under
strong (of the order of 10$^{12}$ G) magnetic fields. A quite evident
manifestation of such transfers are cyclotron resonance scattering
features (CRSFs) observed in the spectra of many X-ray sources --- the
first being observed in Her X-1 \citep{tru78}. CRSFs are caused by
resonant scatterings between the Landau levels of electrons under the
strong magnetic field near the surface of neutron star. The cyclotron
resonance energy, $E_a$, is strongly related to the magnetic field
strength:
\begin{equation}
E_a = 11.6 B_{12} \cdot (1+z_g)^{-1} \, \textrm{keV},
\end{equation}
where $B_{12}$ is the magnetic field strength in units of $10^{12}$ G,
and $z_g$ represents the gravitational red-shift at the resonance
point. To date, CRSFs have been detected from 16 accreting binary
pulsars and measured their magnetic field strength
\citep{miha95,cob02}. However, the formation mechanism of CRSFs is not
still fully understood due to the complex physics of resonant
scattering and their continuum. Thus, the comparison of the CRSFs
parameters observed in different sources, at different luminosity and
spectral states is important to gain information on the physical
processes taking place.

The Japanese X-ray satellite {\it{Suzaku}}, with its high sensitivity
hard X-ray detector \citep[HXD;][]{taka07}, has discovered CRSFs from GX 304-1
\citep{yama11}, successfully detected CRSFs from A0535+262
\citep{tera06} in its lowest luminosity epochs, and performed detailed
studies of the CRSFs of Her X-1 \citep{eno08}, 4U 1907+09
\citep{riv2010} and 1A 1118$-$61 \citep{such11}.

4U 1626$-$67 was observed by {\it{Suzaku}} on 2006 March. The source
is a 7.7~s pulsar in a low-mass X-ray binary system with an orbital
period of 42 min \citep{mid81, cha98}. The X-ray spectrum shows
low-energy emission lines from Ne and O \citep{ang95,sch01, kra07} and
a feature at $\sim 37$ keV \citep{orl98}. The 37 keV absorption
feature was discovered by {\it{BeppoSAX}} and interpreted as the
fundamental CRSF. The 10--60 keV source flux is $4.4 \times 10^{-10}$
erg cm$^{-2}$ s$^{-1}$ and the resonance energy corresponds to a
magnetic field strength of $\sim 3 \times 10^{12}$ G
\citep{orl98}. The broadband X-ray continuum exhibits a strong phase
dependence as reported by {\it{HEAO-1}} \citep{pra79}, {\it{Tenma}}
\citep{kii86} and {\it{RXTE}} \citep{cob01}. The cyclotron feature and
the phase dependence variability provide information on radiative
transfer in the plasma of the accretion column. In this paper, we used
the March 2006 {\it {Suzaku}} observation of 4U 1626$-$67 to
investigate the characteristics of the $\sim 37$ keV feature and the
continuum in the phase resolved spectra. The paper is organized as
follows: we describe the observations from {\it {Suzaku}} and the data
reduction procedure in \S2, show the results of the timing and
spectral analyses in \S3, and finally provide a discussion and summary
in \S4.

\section{Observations and data reduction}

\subsection{ {\it{Suzaku}} observation of 4U 1626$-$67}
 {\it{Suzaku}} is the fifth Japanese X-ray satellite \citep{mitsu07}
 to be launched and is equipped with two different instruments. One
 instrument is an X-ray imaging spectrometer \citep[XIS;][]{koya07}, a
 charge-coupled device camera at the focus of the X-ray telescope
 \citep[XRT;][]{ser07}, covering the 0.2--12 keV energy range.  There are four
 XIS units; three of them (XIS-0, XIS-2 and XIS-3) are
 front-illuminated (FI) while the other one (XIS-1) is back-illuminated
 (BI). The other instrument is a hard X-ray detector \citep[HXD;][]{taka07},
 which consists of PIN silicon diodes (HXD-PIN; 10--70 keV) and
 Gd$_2$SiO$_5$Ce (GSO; 50--600 keV) scintillators. 4U1626$-$67 was
 observed by {\it{Suzaku}} on 2006 March 9 UT 01:18 through March 11
 19:38 at ``XIS nominal'' pointing position. The XIS operated in
 standard clocking mode and included the ``1/8 window'' option in
 order to have a time resolution of 1 s, without a charge injection
 function.
 
\subsection{Data reduction procedure}
We used data produced by {\it{Suzaku}} pipeline processing
version 2.0.6.13 with the calibration versions hxd-20090515,
xis-20090403 and xrt-20080709, and the software version HEADAS 6.5.1.

XIS and HXD events were screened by using standard criteria. We
discarded events collected when the earth's limb was less than
5$^\circ$, when the earth's day--night boundary angle from the line of
sight was less than 25$^\circ$, or when the spacecraft was in an orbit
phase within 436 s (for the XIS) or 500 s (for the HXD) of the South
Atlantic Anomaly ingress/egress. For XIS-1, we also excluded events in
the time interval between 2006 March 10 UT 06:58:30 and 07:20:54,
because the cooling system was not properly working during this
period. We accepted only XIS events with standard grades (0, 2, 3, 4
and 6) in the analysis.

XIS spectra and light curves were extracted from a circle with a
radius of 4.3 arcmin centered at the source. Spectra from the three FI
XIS sensors were summed together and the XIS intensity in the
0.4--10.0 keV energy band was $\sim 5$ c s$^{-1}$ for each sensor. We
accumulated the XIS background spectra from a source-free region
having the same exposure area as that of the corresponding
sensors. The derived background spectra, exhibited an intensity of
$\sim 0.02$ c s$^{-1}$ for each sensor in the 0.4--10.0 keV range, and
is considered negligible.

HXD screened events were used to obtain spectra and light curves for
the PIN and GSO sensors. The HXD-PIN simulated non X-ray background
(NXB) model was provided by the {\it{Suzaku}} HXD team, and we used
the METHOD = LCFITDT and METHODV = 2.0 datasets \citep{fuka09}. The
systematic uncertainty of the NXB model is estimated to be 2.31\% for
a 10 ks exposure in the 15--40 keV energy band
\citep{fuka09}. Hereinafter, we assume 3\% systematic errors in the
HXD-PIN analyses, and the cosmic X-ray background (CXB) spectra were
subtracted by using data acquired by {\it{HEAO-1}}
\citep{bol87}. However, the source was not detected by the HXD-GSO.
The NXB-subtracted HXD-GSO light curve and spectrum exceeded the
HXD-GSO NXB by $\sim 1$\%, and the latter were within typical
systematic uncertainties \citep{fuka09}. Net exposures were 102.7 ks
for XIS-0, 2 and 3, 101.4 ks for XIS-1 and 87.6 ks for the HXD.

\section{Analysis and results}
\begin{figure}
\begin{center}
 \resizebox{1.0\columnwidth}{!}
  {\includegraphics[width=8.0cm,height=9.5cm,clip,angle=270]{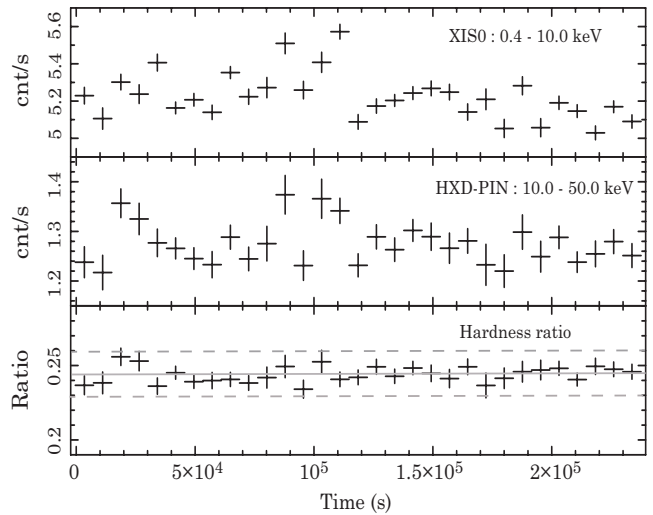}}
 \caption{Energy-resolved light curves of 4U 1626$-$67 observed with {\it{Suzaku}}. Upper and middle panels show background-subtracted light curves obtained with XIS-0 in the 0.4--10.0 keV range and with the HXD in the 10--50 keV range, respectively. The bottom panel shows the hardness ratio of the above two panels.The gray solid line shows the average and the gray dotted lines show the $\pm$3$\sigma$ interval.}
 \label{fig:a}
\end{center}
\end{figure}
\subsection{Timing analysis}
For the arrival time of each XIS and HXD-PIN event, barycenter
correction was performed with the tool aebarycen \citep{tera08}. We
surveyed the HXD-PIN data by using a folding technique to obtain a
best spin period of $P_{(\rm Suzaku)} = 7.67795(9)$ s. This period is
consistent with the spin-down trend that started to be observed from
1990 onwards, and that is calculated by using the spin-down rate of
$\dot{\nu} \sim -4.8 \times 10^{-13}$ Hz s$^{-1}$ measured with the
Swift satellite over the period 2004--2007 \citep{cha97,cam10}.

XIS and HXD background-subtracted light curves of 4U 1626$-$67 are
shown in Figure 1. The light curves are plotted using a time bin of
7678 s; a multiple of $P_{(\rm Suzaku)}$. The XIS and HXD light curves
have an intensity variation of less than 7\%, and the hardness ratio
between them is approximately constant; the reduced chi-squared
($\chi^2$) value is about 1 when fitting a constant model to the
hardness ratio. The average and their 3 $\sigma$ level are shown in
the Figure 1. As a results, we confirmed that there is no significant
spectral variation. Thus, X-ray spectra were extracted by using data over the
entire observation.

\begin{figure}
\begin{center}
 \resizebox{1.0\columnwidth}{!}
  {\includegraphics[width=6cm,height=6cm,keepaspectratio,clip]{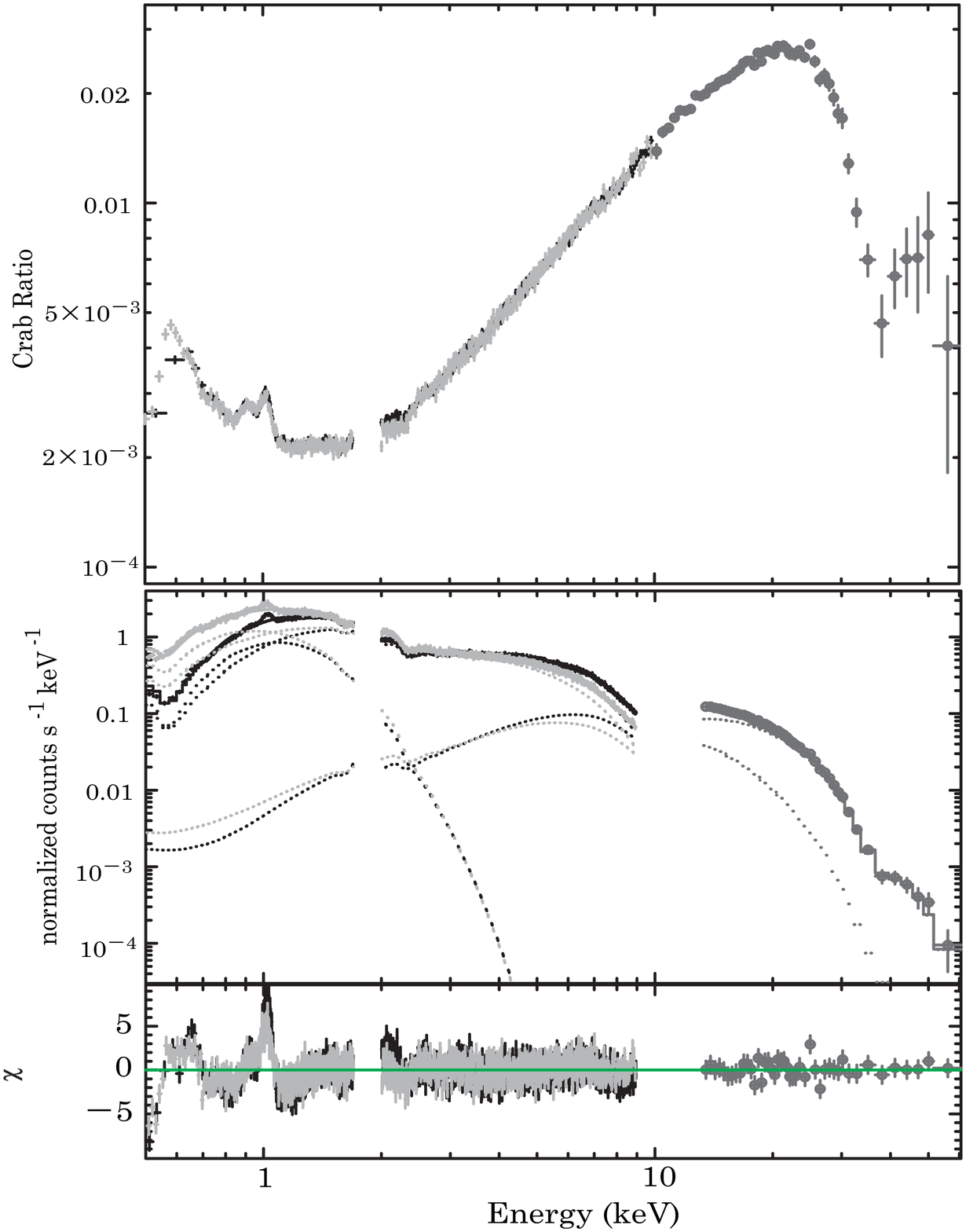}}
 \caption{Phase-averaged background-subtracted spectra of 4U 1626$-$67
   obtained with {\it{Suzaku}} XIS (black crosses denote XIS-FI data
   and light gray crosses denote XIS-BI data) and HXD (denoted by
   circles). Error bars are only for statistical predictions. The top
   panel shows the raw count rate energy spectra divided by those of
   Crab Nebula. The middle panel shows the spectra in counts/s
   keV$^{-1}$ along with the best-fit models (dotted lines) a
   blackbody, NPEX with Gaussian-absorption model and photoelectric
   absorption (see text for details). Plots are presented without
   removal of instrumental responses. The bottom panel shows the
   residuals between the actual data and the best-fit model.}
 \label{fig:d}
\end{center}
\end{figure}
\subsection{Phase-averaged spectrum}

The top panel of Figure 2 shows the broadband 4U 1626$-$67
phase-averaged raw count rate spectrum normalized by a Crab Nebula
spectrum that was observed for calibration purposes. XIS source data
were normalized by using a canonical model of the Crab Nebula
spectrum, since spectra from the Crab Nebula suffer from pile-up. HXD
data were normalized by a HXD Crab Nebula spectrum observed on 2006
March 30 taken for the same observation parameters---position of the
source with respect to the instrument FoV, grade criteria and PIN
threshold file (ae\_hxd\_pinthr\_20060727.fits)---as for 4U
1626$-$67. The 4U 1626$-$67 source intensity is about 20 mCrab at 20
keV, and the normalized spectrum is characterized by three main
features: a soft X-ray excess below $\sim 3$ keV, as reported by
several authors \citep{ang95,sch01,kra07}; a power-law continuum and a
dip feature at $\sim$35 keV.

To model the phase-averaged spectrum, we fit the data with a $\sim
0.24$ keV blackbody for the soft excess, a negative and positive
power-law times exponential (NPEX) model \citep{miha95, maki99} and a
Gaussian-absorption (GABS) model \citep{soo90} for the continuum above
3 keV and a standard photoelectric absorption. The NPEX model
\begin{equation}
{f(E) = (A_{n}E^{-\alpha}+A_{p}E^{+\beta}) \exp \left( -\frac{E}{kT}\right), 
}
\end{equation}
where $E$ is the energy of an incident photon and $kT$ is a plasma
temperature (also called the cutoff energy), has two power law
components with indices $\alpha$ and $\beta$. The negative $\alpha$
term describes the spectrum at low energy, and the positive $\beta$
term, which dominates as the energy increases, simulates the Wien peak
in the unsaturated Comptonization regime when $\beta=2.0$.

\begin{figure}
\begin{center}
 \resizebox{1.0\columnwidth}{!}
  {\includegraphics[width=6cm,height=6cm,keepaspectratio,clip]{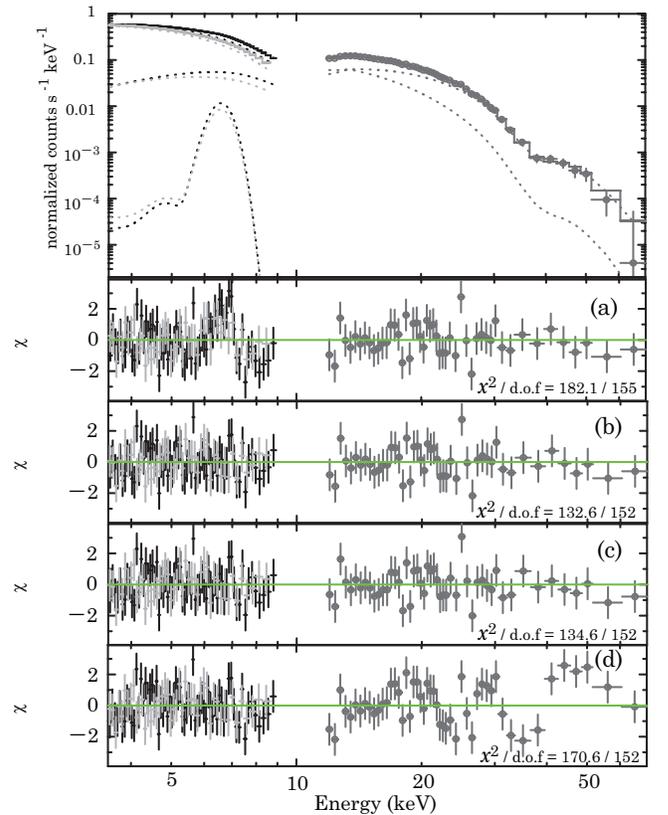}}
 \caption{The top panel shows the phase-averaged and
   background-subtracted spectra above 3.5 keV of 4U 1626$-$67 obtained
   with {\it{Suzaku}} XIS (FI: black crosses, BI: light
   gray crosses) and HXD (gray circles). The lower
   four panels show the residuals in units of standard deviations, $\chi$, from fitting by: (a) the NPEX with GABS
   model without Fe-line emission model; (b) the NPEX with
   GABS model and Fe-line emission model; (c) the NPEX with CYAB
   and Fe-line emission model and (d) the ECUT with GABS model and Fe-line emission model.}
 \label{fig:e}
\end{center}
\end{figure}
\begin{deluxetable*}{cccccccccccc}
\tabletypesize{\scriptsize}
\tablecaption{\label{tab:table_temp}Best-fit parameters for phase-averaged spectra of 4U 1626$-$67 (d.o.f. = degrees of freedom)}
\tablewidth{0pt}
\tablehead{
&\colhead{Parameter} & & \colhead{NPEX}\tablenotemark{a} & &\colhead{NPEX with GABS}\tablenotemark{a}&
&\colhead{NPEX with GABS} &&\colhead{NPEX with CYAB} & &
\colhead{ECUT with GABS}}
\setlength{\tabcolsep}{1.5pt}
\startdata

&$\alpha$& &$0.31^{+0.03}_{-0.03}$  & &$0.26^{+0.02}_{-0.02}$ & &$0.29^{+0.02}_{-0.02}$ & &$0.32^{+0.03}_{-0.03}$ & &$0.65^{+0.06}_{-0.07}$ \\
&$kT $(keV) & &$5.3^{+0.1}_{-0.1}$  & &$7.0^{+0.5}_{-0.4}$ & &$6.7^{+0.2}_{-0.3}$ & &$7.2^{+1.1}_{-0.4}$ & &---   \\
&$E_c$(keV)& &---&&---&&--- & &--- & &$24.7^{+0.43}_{-0.53}$ \\
&$E_f$(keV)& &---&&---&&--- & &--- & &$7.08^{+1.14}_{-0.81}$ \\
&$A_n(\times 10^{-3})$\tablenotemark{b} & &$8.1^{+0.4}_{-0.3}$ & &$6.8^{+0.2}_{-0.2}$ & &$7.2^{+0.3}_{-0.3}$& &$7.3^{+0.1}_{-0.3}$ & &---     \\
&$A_p(\times 10^{-5})$\tablenotemark{b} & &$4.6^{+0.3}_{-0.3}$& &$1.6^{+0.5}_{-0.4}$ & &$1.8^{+0.5}_{-0.5}$& &$1.6^{+0.4}_{-0.6}$ & &---     \\
&$A(\times 10^{-3})$\tablenotemark{c} & &--- & &--- & &--- & &--- & &$10.0^{+1.39}_{-0.75}$   \\
&$E_{Fe}$(keV)& &--- & &--- & &$6.62^{+0.10}_{-0.11}$ & &$6.60^{+0.10}_{-0.11}$ & & $6.60 $(fix) \\
&$\sigma_{Fe}$(keV)& &--- & &--- & &$0.37^{+0.13}_{-0.10}$ & &$0.39^{+0.13}_{-0.10}$ & & $0.37^{+0.10}_{-0.10}$ \\
&$EW_{Fe}$[eV] & &--- & &--- & &$33.3^{+16.1}_{-16.5}$ & &$35.2^{+27.7}_{-11.0}$ & &$31.3^{+3.2}_{-7.6}$\\
&$E_a$(keV) & &--- & &$37.6^{+1.02}_{-0.89}$ & &$37.4^{+1.0}_{-0.9}$ & &$35.7^{+0.65}_{-0.55}$ & &$38.4^{+6.47}_{-4.80}$   \\
&$\sigma_{CRSF} $(keV) & &--- & &$5.11^{+0.79}_{-0.68}$ & &$4.99^{+0.71}_{-0.65}$ & &$5.56^{+1.77}_{-1.08}$ & &$>16.5$ \\
&$\tau $ & &--- & &$20.4^{+4.87}_{-4.20}$ &  &$19.0^{+5.0}_{-3.6}$ & &$25.1^{+3.5}_{-2.3}$ & &$79.1^{+9.74}_{-4.59}$   \\
&$\chi ^{2}$ / d.o.f& &572.6/158 & &182.1/155 & &132.6/152 & &134.6/152 & &170.6/152

\enddata
\tablenotetext{a}{Without Fe-emission line model.}
\tablenotetext{b}{Referring to equation (1), and defined at 1 keV in units of photons keV$^{-1}$ cm$^{-2}$ s$^{-1}$.}
\tablenotetext{c}{Normalization of the power-law. Defined at 1 keV in units of photons keV$^{-1}$ cm$^{-2}$ s$^{-1}$.}
\end{deluxetable*}
The best-fit model and its residuals are plotted in the lower panel of
Figure 2. Above 3.5 keV only 1\% of the NPEX negative term is
contributed by the blackbody component. Therefore, hereinafter, we
ignore XIS data below 3.5 keV and exclude the blackbody component from
the fit.

To evaluate the statistical significance of the CRSF in the phase
averaged spectrum, first, we fit the spectrum with only NPEX continuum
or NPEX $\times$ GABS model. The residuals of the NPEX with GABS model
for energy bands above 3.5 keV are shown in Figure 3(a), and the
best-fit parameters are listed in Table 1. Second, we tested the
significance using the F-test of \citet{pre07} routine. In this case,
the F statistical value is defined as
\begin{equation}
{F = \frac{\chi_{1}^{2}/\nu_{1}}{\chi_{2}^{2}/\nu_{2}}, 
}
\end{equation}
where the $\chi_{1}^{2}$ and $\chi_{2}^{2}$ are statistic chi-squared,
$\nu_{1}$ and $\nu_{2}$ are degrees of freedom corresponding to the
results of fittings using model 1 and model 2 (In this case model 1 is
only NPEX continuum and model 2 is NPEX with multiplicative
component), respectively. The derived $F$ statistic value is 3.1 which
indicates that the probability of chance improvement of the $\chi^{2}$
is $3.7 \times 10^{-12}$. Thus, the CRSF is statistically significant
feature in the phase averaged spectrum. We also found that the
residuals in Figure 3(a) have a complex structure near the Fe-emission
line energy around 6.6 keV. The fit improves by the addition of a
Gaussian Fe-line model with an energy of 6.62 keV, a width
($\sigma_{Fe}$) of 0.37 keV and an equivalent width (EW) of 33.3 eV
(Figure 3(b) and Table 1). These values are statistically consistent
with the {\it{ASCA}} EW upper limit of $< 33$ eV for an Fe-emission
line in the 6.4--6.9 keV range \citep{ang95}. The broad width may
indicate the presence an Fe-line complex. If we use three narrow
Gaussian Fe-lines, the Fe-line energies are $6.37^{+0.07}_{-0.08}$,
$6.61^{+0.05}_{-0.07}$ and $6.92^{+0.04}_{-0.04}$ keV with
corresponding EWs of $< 11.1$, $< 12.2$ and $< 17.4$ eV, although the
fit is not improved significantly and we would not be distinguish
whether the Fe-emission line structure is made by three lines or not,
due to the energy resolution of XIS ($\sim$150 eV). The X-ray
unabsorbed fluxes in the 3.5--10 keV and 10--60 keV energy bands are
$1.27^{+0.01}_{-0.16} \times10^{-10}$ erg cm$^{-2}$ s$^{-1}$ and
$4.02^{+0.46}_{-0.88} \times10^{-10}$ erg cm$^{-2}$ s$^{-1}$,
respectively, and the hard X-ray flux is consistent with the
{\it{BeppoSAX}} observation \citep{orl98}. Exchanging the GABS model
with a cyclotron absorption model \citep[CYAB;][]{miha90} also fits the data
well (Figure 3(c) and Table 1).

If we replace the NPEX model by an exponential cutoff power-law model
\citep[ECUT, which is the \texttt{powerlaw*highecut} model in
  Xspec;][]{whi83} and fit the data with a GABS model, we obtain an
acceptable $\chi^2$ value (see Table 1), but substantial residual
structures are seen above 20 keV (Figure 3(d)). Although the CRSF
energy agrees with {\it{BeppoSAX}} observations \citep{orl98}, the
best-fit continuum model is inconsistent with {\it{BeppoSAX}} results.

\subsection{Phase-resolved spectra}

Phase-resolved spectra acquired by {\it{HEAO-1}}, {\it{Tenma}} and
{\it{RXTE}} \citep{pra79,kii86,cob01}, show a strong phase dependence
in the continuum emission of 4U 1626$-$67. {\it{Suzaku}} XIS and HXD
folded light curves also indicate changes across the pulse (Figure
4). Specifically, while the 10--30 keV / 3.5--10 keV ratio maintains
the characteristics of the individual light curves, the 30--50 keV /
10--30 keV ratio indicates spectral differences between phases
0.0--0.3 and the other phases.
To investigate spectral changes across the pulse, we divided the
events into eight phase-resolved spectra (Figure 5) and normalized
these spectra by the Crab Nebula spectrum by using the method
described in \S3.2 (Figure 6). The Crab Ratios show that the $\sim 20$
keV peak is more prominent at higher pulse flux. Additionally, the
Ratios flatten at lower pulse flux and the energy of resonance feature
changes. The resonance features are particularly outstanding in the
dim phase spectrum ($\phi = 0.125$--0.250), where the hardness ratio
(Figure 4) shows the large change in these features.
\begin{figure*}
\begin{center}
 \resizebox{1.0\columnwidth}{!}
  {\includegraphics[width=8.0cm,height=9.5cm,clip]{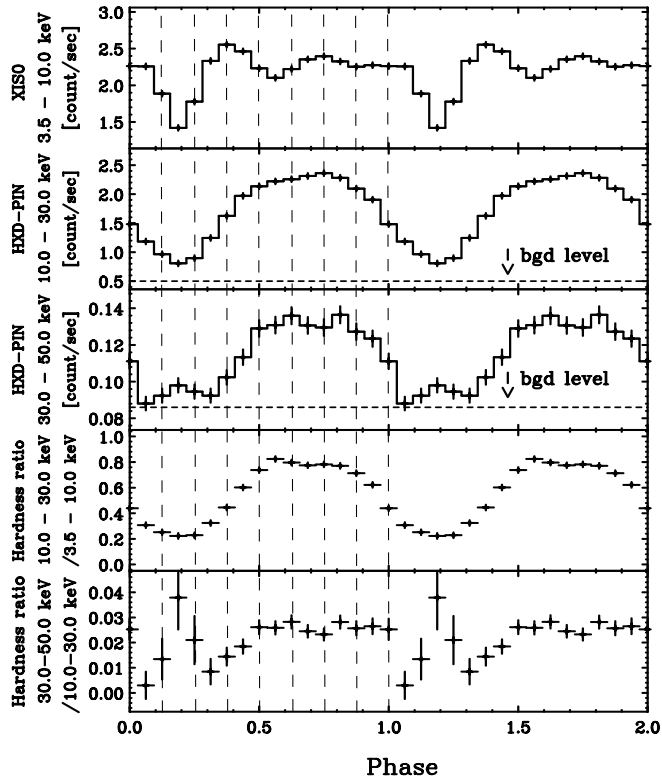}}
 \caption{Energy-resolved and background-inclusive pulse profiles of
   4U 1626$-$67. Data were folded at $P_{(\rm Suzaku)}=7.67795$
   s. Phase 0 corresponds to MJD 53803.000092755. The average
   background count rates of 3.5--10.0 keV, 10.0--30.0 keV and
   30.0--50.0 keV energy bands are 0.016 counts/s, 0.49 counts/s and
   0.086 counts/s, respectively.}
 \label{fig:c}
\end{center}
\end{figure*}


\begin{deluxetable*}{cccccccc}
\tabletypesize{\scriptsize}
\tablecaption{\label{tab:table_temp}NPEX continuum-only best-fit parameters for phase resolved spectra of 4U 1626$-$67.}
\tablewidth{0pt}
\tablehead{
\colhead{pulse} & \colhead{$\alpha$} &\colhead{kT}
&\colhead{$A_n$}\tablenotemark{a} &\colhead{$A_p$}\tablenotemark{a}
&\colhead{EW$_{Fe}$} \tablenotemark{b} & \colhead{$\chi ^{2}$ / d.o.f}\\
\colhead{phase} &  &\colhead{(keV)}
&\colhead{$(\times 10^{-3})$}&\colhead{$(\times 10^{-5})$}
&\colhead{(eV)} & }
\setlength{\tabcolsep}{0.5pt}
\startdata

0.000--0.125 & $-0.02^{+0.05}_{-0.06}$  &  $4.4^{+0.4}_{-0.3}$  & $6.7^{+0.8}_{-0.7}$
& $3.2^{+2.3}_{-1.9}$ &$27^{+38}_{-2}$& 81.2 / 81\\

0.125--0.250 & $0.15^{+0.17}_{-0.11}$  &  $5.1^{+0.7}_{-1.1}$  & $5.8^{+0.7}_{-0.5}$ & $< 1.2  $ &$81^{+28}_{-12}$&  69.0 / 63\\

0.125--0.250 & $0.27^{+0.02}_{-0.02}$  &  $6.1^{+0.5}_{-0.4}$  & $6.1^{+0.2}_{-0.2}$ & 0 (fixed)  &$84^{+13}_{-53}$&  71.8 / 64\\

0.250--0.375 & $0.10^{+0.07}_{-0.06}$  &  $4.3^{+0.2}_{-0.3}$  & $8.04^{+1.3}_{-0.7}$
& $4.9^{+2.5}_{-1.7}$ &$46^{+31}_{-8}$& 93.4 / 97\\

0.375--0.500 & $0.51^{+0.11}_{-0.06}$  & $5.0^{+0.1}_{-0.1}$  & $11.8^{+1.9}_{-0.8}$
& $7.6^{+1.1}_{-0.7}$ &$21^{+16}_{-9}$& 137.3 / 103\\

0.500--0.625 & $0.56^{+0.10}_{-0.10}$  & $5.3^{+0.1}_{-0.1}$  & $10.4^{+1.6}_{-1.3}$ & $7.2^{+0.6}_{-0.6}$ &$< 22$ &297.9 / 129\\
0.625--0.750 & $0.64^{+0.10}_{-0.09}$  & $5.2^{+0.1}_{-0.1}$  & $11.8^{+1.7}_{-1.5}$ & $8.4^{+0.7}_{-0.7}$ & $35^{+5}_{-10}$ &259.0 / 108\\
0.750--0.875 & $0.53^{+0.09}_{-0.09}$  & $5.3^{+0.1}_{-0.1}$  & $9.8^{+1.4}_{-1.2}$ & $7.5^{+0.7}_{-0.7}$ & $35^{+17}_{-15}$ & 247.2 / 91\\
0.875--1.000 & $0.45^{+0.07}_{-0.07}$  & $5.3^{+0.2}_{-0.1}$  & $9.9^{+1.08}_{-0.9}$ & $5.7^{+0.7}_{0.7}$ & $31^{+33}_{-27}$ & 117.5/ 89\\
\enddata
\tablenotetext{a}{Referring to equation (1), and defined at 1 keV in
  units of photons keV$^{-1}$ cm$^{-2}$ s$^{-1}$.}
\tablenotetext{b}{The energy and the width of the Fe-line are fixed at
  6.62 keV and $\sigma_{Fe}=0.37$ keV, respectively.}
\end{deluxetable*}

\begin{deluxetable*}{cccccccccccc}
\tabletypesize{\scriptsize}
\tablecaption{\label{tab:table_temp}Best-fit parameters for phase resolved spectra of 4U 1626$-$67, when fitting by NPEX with GABS model}
\tablewidth{0pt}
\tablehead{
\colhead{pulse} & \colhead{$\alpha$} &\colhead{kT}
&\colhead{$A_n$}\tablenotemark{a} &\colhead{$A_p$}\tablenotemark{a} &
\colhead{$E_a$}& \colhead{$\sigma_{CRSF}$} & \colhead{$\tau$}
&\colhead{EW$_{Fe}$} \tablenotemark{b} & \colhead{$\chi ^{2}$ / d.o.f}
& \colhead{$F$/PCI} \tablenotemark{c}\\
\colhead{phase} &  &\colhead{(keV)}
&\colhead{$(\times 10^{-3})$}&\colhead{$(\times 10^{-5})$} &
\colhead{(keV)}& \colhead{(keV)} &
&\colhead{(eV)} & & }
\setlength{\tabcolsep}{1.0pt}
\startdata

0.000--0.125 & $0.05^{+0.18}_{-0.07}$  &  $5.4^{+1.8}_{-0.8}$  & $6.5^{+0.5}_{-0.4}$
& $< 2.4 $ &  $31.3^{+2.3}_{-1.6}$ & $0.5^{+2.5}_{-0.3}$ & $> 5 $ &$ <57$ & 66.9 / 78 & 1.169/$2.45\times 10^{-1}$\\

0.125--0.250 & $0.29^{+0.26}_{-0.40}$  &  $ 11.2^{+4.4}_{-1.8} $ & $7.8^{+10.8}_{-1.7}$ & 0 (fixed)  & $26.3^{+5.6}_{-4.1}$ & $> 9.1$ & $ 66^{+94}_{-42} $ & $69^{+46}_{-35}$ & 52.0 / 61 &1.316/$1.41\times 10^{-1}$\\

0.250--0.375 & $0.21^{+0.29}_{-0.22}$  &  $6.5^{+2.6}_{-3.0}$  & $7.5^{+1.0}_{-0.5}$
& $< 4.6  $ & $33.5^{+8.1}_{-3.2}$ & $4.5^{+4.7}_{-1.2}$ & $15^{+47}_{-8}$ &$43^{+23}_{-28}$ & 85.3 / 94 &1.061/$3.87\times 10^{-1}$\\

0.375--0.500 & $0.40^{+0.07}_{-0.07}$  & $5.6^{+0.6}_{-0.2}$  & $9.8^{+1.2}_{-1.9}$
& $5.0^{+1.1}_{-2.0}$ & $37.1^{+3.8}_{-0.9}$ & $4.0^{+2.8}_{-1.5}$ & $14^{+15}_{-6.0}$ &$24^{+37}_{-7}$ & 91.9 / 100&1.451/$3.14\times 10^{-2}$\\
0.500--0.625 & $0.33^{+0.07}_{-0.07}$  & $6.6^{+0.5}_{-0.4}$  & $6.8^{+0.9}_{-0.8}$
& $3.4^{+0.9}_{-1.0}$ & $37.2^{+2.2}_{-1.2}$ & $3.9^{+1.2}_{-1.1}$ & $22^{+11}_{-6}$ & $< 45$ & 133.6 / 126 &2.178/$7.62\times 10^{-6}$\\
0.625--0.750 & $0.38^{+0.02}_{-0.03}$  & $6.4^{+0.1}_{-0.1}$  & $8.0^{+0.1}_{-0.1}$
& $4.2^{+0.5}_{-0.1}$ & $36.8^{+1.0}_{-0.9}$ & $4.0^{+1.1}_{-0.8}$ & $20^{+4}_{-3}$ &$39^{+6}_{-19}$ & 83.6 / 105 &3.012/$1.63\times 10^{-8}$\\
0.750--0.875 & $0.27^{+0.05}_{-0.06}$  & $7.1^{+0.8}_{-0.5}$  & $6.7^{+0.7}_{-0.7}$
& $2.6^{+1.0}_{-1.2}$ & $39.0^{+2.5}_{-1.9}$ & $5.5^{+1.4}_{-1.2}$ & $29^{+29}_{-9}$ & $38^{+26}_{-22}$&108.7 / 88 &2.199/$1.25\times 10^{-4}$\\
0.875--1.000 & $0.37^{+0.15}_{-0.07}$  & $8.0^{+20.0}_{-2.5}$  & $7.6^{+0.8}_{-0.5}$
& $1.4^{+3.0}_{-0.7}$ & $43.3^{+9.4}_{-8.5}$ & $10.0^{+4.6}_{-3.2}$ & $> 10$ &$30^{+28}_{-25}$& 90.4/ 86 &1.256/$1.45\times 10^{-1}$\\
\enddata
\tablenotetext{a}{Referring to equation (1), and defined at 1 keV in units of photons keV$^{-1}$ cm$^{-2}$ s$^{-1}$.}
\tablenotetext{b}{The energy and the width of the Fe-line are fixed at 6.62 keV and $\sigma_{Fe}=0.37$ keV, respectively.}
\tablenotetext{c}{Referring to equation (3), and PCI is probability of chance improvement of the $\chi^2$ by adding a new component.}
\end{deluxetable*}


\begin{figure}
\begin{center}
 \resizebox{1.0\columnwidth}{!}
  {\includegraphics[width=17cm,height=25cm]{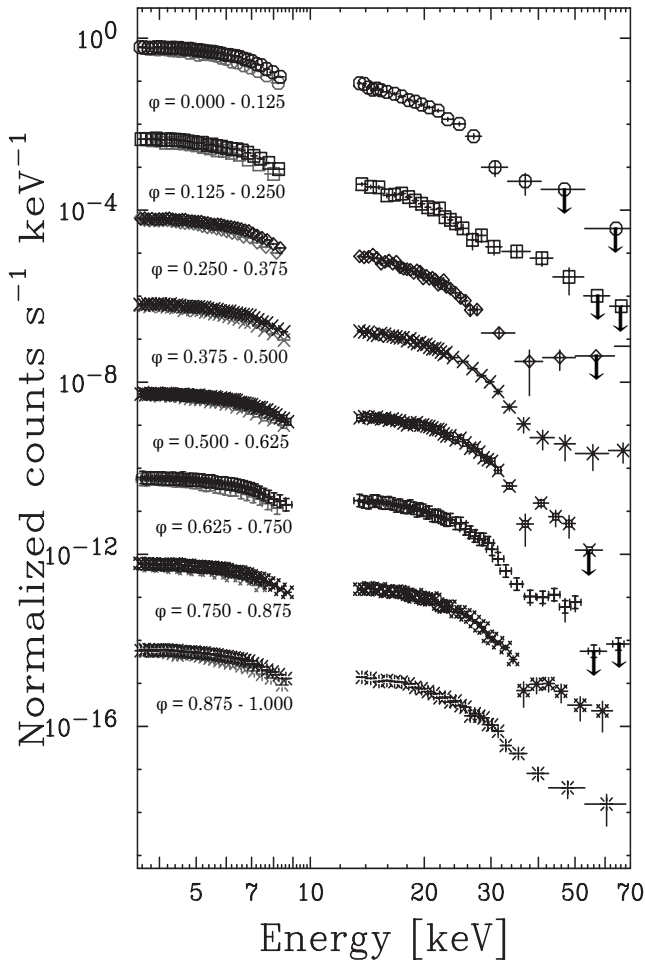}}
 \caption{Phase-resolved and background-subtracted XIS (FI: black
   symbols below 10 keV, BI: gray symbols) and HXD (black symbols over
   10 keV) spectra of 4U 1626$-$67 for phases $\phi = 0.000$--0.125,
   0.125--0.250, 0.250--0.375, 0.375--0.500, 0.500--0.625,
   0.625--0.750, 0.750--0.875 and 0.875--1.000 from top to bottom,
   offset by multiples of $10^{-2}$. Arrows denote the upper limit.}
 \label{fig:d}
\end{center}
\end{figure}

\begin{figure}
\begin{center}
 \resizebox{1.0\columnwidth}{!}
  {\includegraphics[width=17cm,height=25cm]{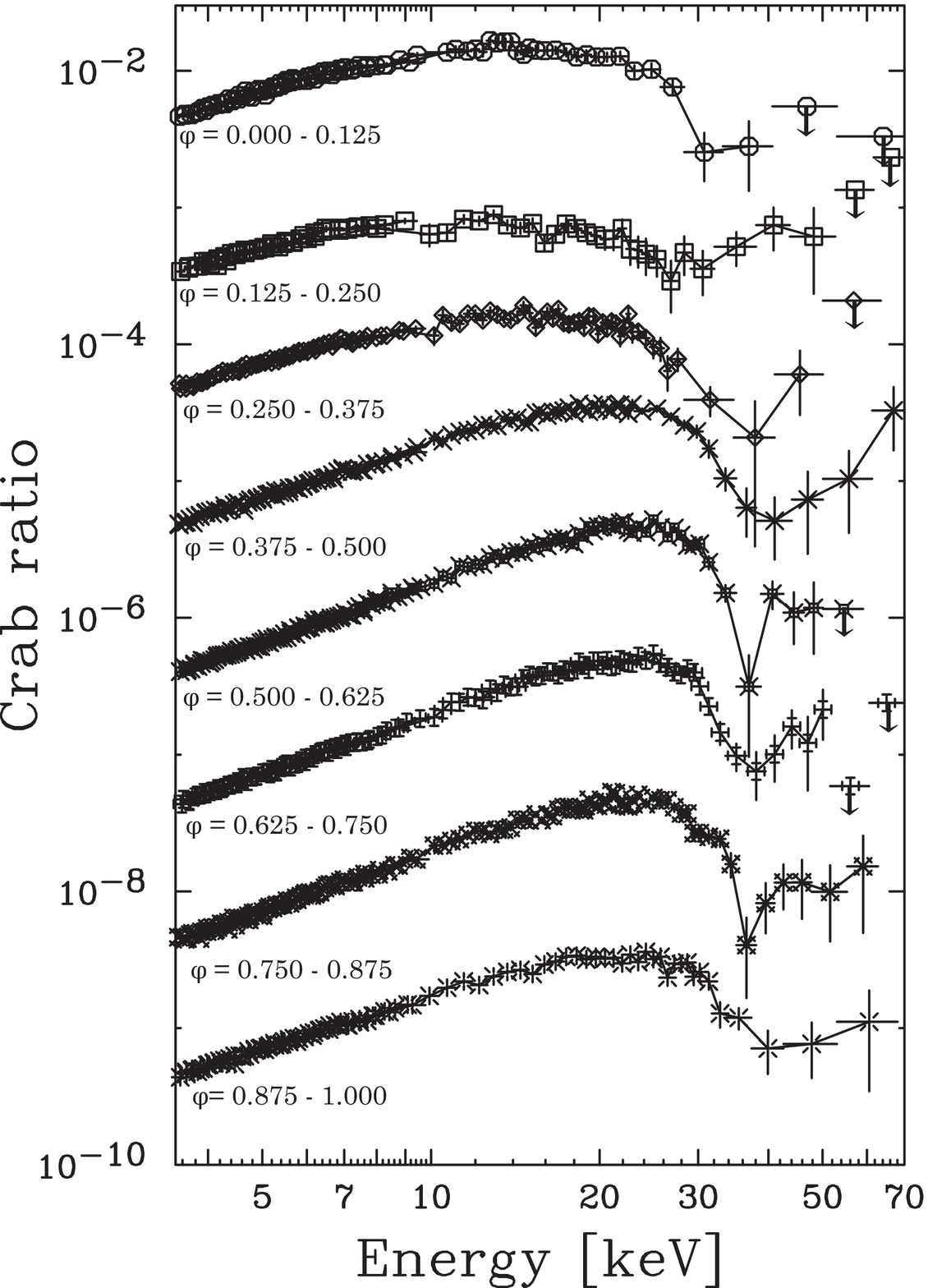}}
 \caption{Ratios between the phase-resolved and background-subtracted XIS
   and HXD spectra of 4U 1626$-$67 and those of the Crab Nebula. The XIS
   source data were normalized using the canonical model of the Crab Nebula
   since spectra from the Crab Nebula suffer from pile-up. Phases are
   indicated in the same manner as Figure 5, but are offset by multiples of $10^{-1}$. Arrows denote the upper limit.}
 \label{fig:e}
\end{center}
\end{figure}

We then performed spectral fitting to the phase-resolved spectra with
the NPEX continuum and the Fe-emission line model in order to
establish the pulse-phase dependence of the continuum and the
CRSF. The Fe-line energy and the width of the Gaussian function were
fixed at values obtained from the averaged spectrum. The parameters
from the spectral fitting are listed in Table 2, and the ratios
between the original data and the best-fit model are shown in Figure
7. We found that the ratio for the dim phase ($\phi = 0.125$--0.250)
spectrum exhibits a monotone increase above 30 keV, and the residuals
show evidence for the existence of an emission feature rather than an
absorption (Figure 8(a)). In contrast, the other phases have a
V-shaped ratio caused by a CRSF absorption feature. Moreover, the
fitting results for the dim phase spectrum show that the $A_p$
parameter is unconstrained, strongly suggesting that the continuum
spectrum at the dim phase requires only the upper limit of the
high-energy cutoff component. In other words, the result make the NPEX
model effectively a power-law times exponential cutoff model
(\texttt{cutoffpl} in Xspec). The result is consistent with the trend
that the 20 keV peak in the Crab Ratios is less prominent at lower
pulse flux. We also tried other empirical models applied to accretion
powered pulsar, ECUT and a powerlaw with Fermi-Dirac cutoff
\citep[FDCO;][]{tanaka1986}, and find that their best fit cutoff
energies are consistent with 0. In other words, a power-law times an
exponential cutoff model is the only one able to fit the dim phase
spectrum. For this reason we fit the dim phase spectrum by using the
NPEX model with fixed $A_p = 0$. As we can see in Table 2, the
difference between the $\chi^2$ values for freely selected $A_p$ and
for fixed $A_p = 0$ is negligible. Therefore, we henceforth fix $A_p =
0$ for the NPEX model when fitting the dim phase spectrum.

\begin{figure}
\begin{center}
 \resizebox{1.0\columnwidth}{!}
  {\includegraphics[width=17cm,height=25cm]{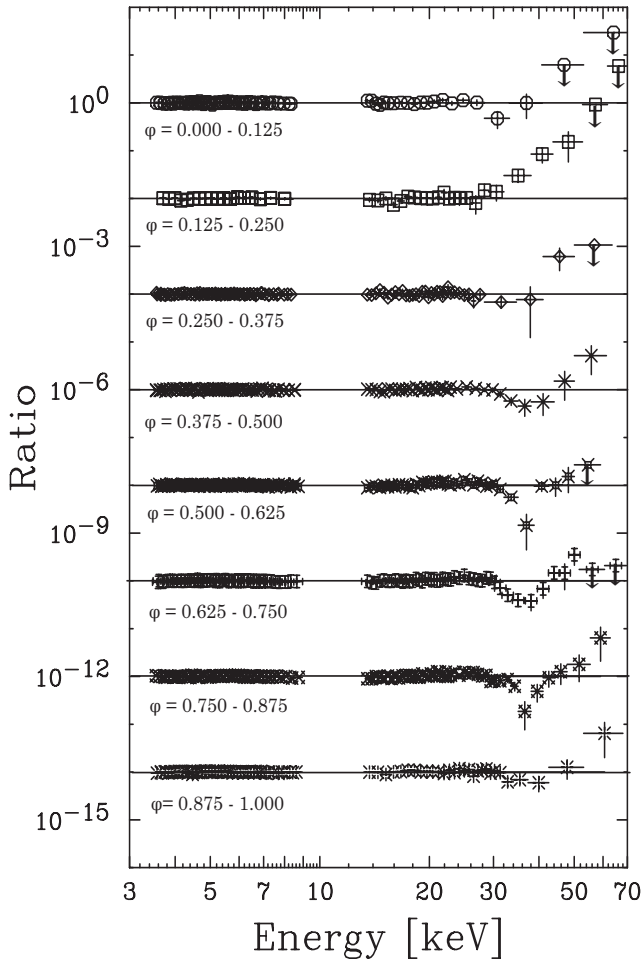}}
 \caption{Ratios between the data and best-fit NPEX-only model for
    each phase-resolved spectra. Phases are indicated in the same manner as Figure 5. Arrows
   denote the upper limit.}
 \label{fig:f}
\end{center}
\end{figure}


\begin{figure}
\begin{center}
 \resizebox{1.0\columnwidth}{!}
{\includegraphics[width=6.0cm,height=8.0cm]{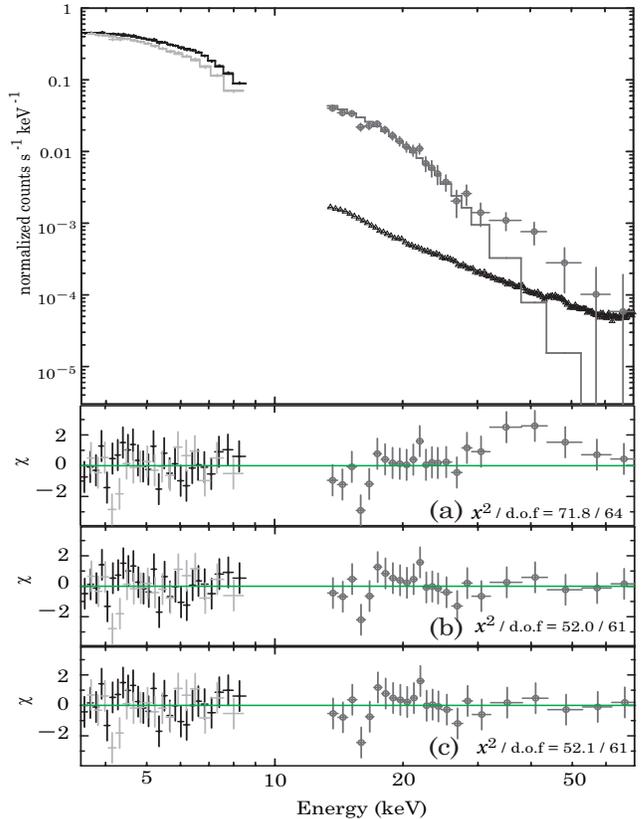}}
 \caption{Upper panel shows spectra in the dim phase ($\phi = 0.125$--0.250) of 4U 1626$-$67 obtained with  {\it{Suzaku}} XIS (FI: black crosses, BI: light
   gray crosses) and HXD (circles). Triangles denote the reproducibility of NXB for HXD-PIN. Histograms in the upper panel are the best-fit by NPEX continuum ($A_p = 0$). The lower three panels show residuals, $\chi$, when fitting by (a) NPEX ($A_p = 0$) continuum-only; (b) NPEX ($A_p = 0$) with GABS model and (c) NPEX ($A_p = 0$) and Gaussian models.}
\label{fig:e}
\end{center}
\end{figure}
\begin{figure*}
\begin{center}
 \resizebox{1.0\columnwidth}{!}
  {\includegraphics[width=17cm,height=25cm]{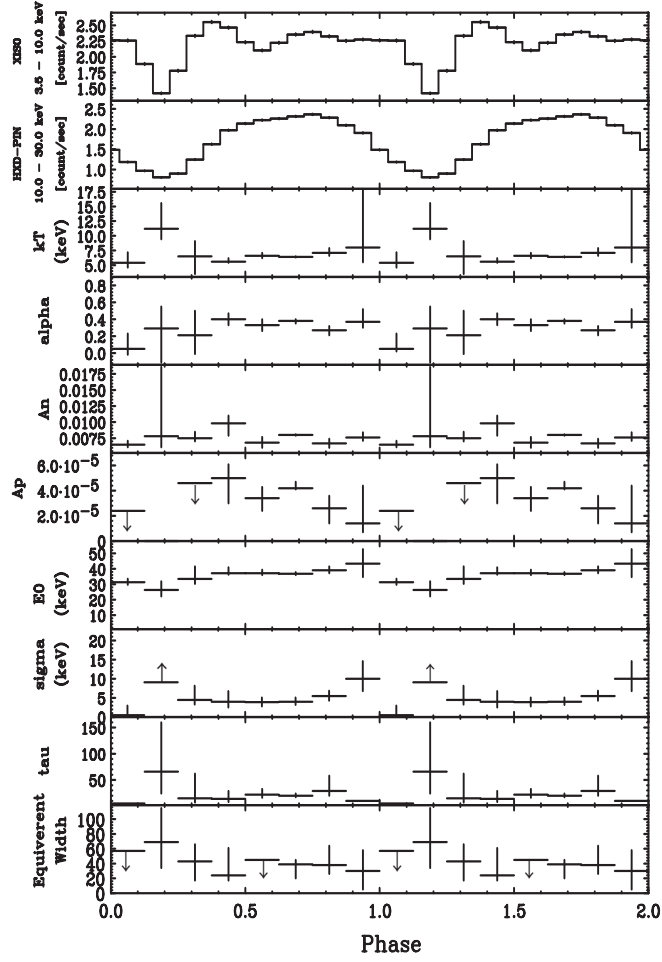}}
 \caption{Fit parameters from as a function of pulse phase for 4U 1626$-$67. Best-fit values by NPEX with GABS model (Table 3) are plotted.}
 \label{fig:f}
\end{center}
\end{figure*}
Next, we included the GABS model and refit the phase-resolved
spectra. The resulting parametrized models fit all of the
phase-resolved spectra well (see Table 3 and Figure 9), and residuals
above 30 keV in the dim phase were eliminated (see Figure 8(b)). We
also evaluate the significance of including a GABS component using the
same method discussed in section 3.1. The derived probabilities are
summarized in Table 3, that shows that the absorption factor is not
statistically significant in the dim phase.  This significance did not
change even when we adopted the CYAB model (as opposed to the GABS
model as the cyclotron feature model) for spectral fitting.
\begin{deluxetable*}{cccccccccccc}
\tabletypesize{\scriptsize}
\tablecaption{\label{tab:table_temp}Best-fit parameters for phase resolved spectra of 4U 1626$-$67, when fitting by NPEX plus Gaussian model.}
\tablewidth{0pt}
\tablehead{
\colhead{pulse} & \colhead{$\alpha$} &\colhead{kT}
&\colhead{$A_n$}\tablenotemark{a} &\colhead{$A_p$}\tablenotemark{a} &
\colhead{$E_a$}& \colhead{$\sigma_{CRSF}$} & \colhead{Norm\tablenotemark{b}}
&\colhead{EW$_{Fe}$} \tablenotemark{c} & \colhead{$\chi ^{2}$ / d.o.f}
& \colhead{$F$/PCI} \tablenotemark{d}\\
\colhead{phase} &  &\colhead{(keV)}
&\colhead{$(\times 10^{-3})$}&\colhead{$(\times 10^{-5})$} &
\colhead{(keV)}& \colhead{(keV)} & \colhead{$(\times 10^{-4})$}
&\colhead{(eV)} & & }
\setlength{\tabcolsep}{1.0pt}
\startdata

0.000--0.125 & $0.04^{+0.19}_{-0.08}$  &  $5.3^{+1.8}_{-0.8}$  & $6.5^{+0.5}_{-0.4}$
& $< 3.1 $ &  $30.8^{+1.8}_{-1.4}$ & $ < 3.4$ & $ -2.3^{+1.1}_{-1.6} $ &$ 27^{+16}_{-10}$ & 66.8 / 78 & 5.617/ 1.54$\times 10^{-3}$\\

0.125--0.250 & $0.15^{+0.14}_{-0.17}$  &  $ 5.4^{+0.4}_{-0.8} $ & $5.5^{+0.8}_{-0.7}$ & $ 0 $(fixed)  & $40.5^{+13.7}_{-14.7}$ & $> 4.5$ & $ 8.8^{+7.2}_{-2.9} $ & $71^{+29}_{-36}$ & 52.1 / 61& 7.688/1.94$\times 10^{-4}$\\

0.250--0.375 & $0.21^{+0.20}_{-0.15}$  &  $6.3^{+2.2}_{-1.9}$  & $7.6^{+0.5}_{-0.6}$
& $< 4.5  $ & $29.9^{+2.7}_{-5.6}$ & $5.4^{+4.6}_{-4.9}$ & $-6.7^{+5.6}_{-17.4}$ &$47^{+17}_{-17}$ & 85.2 / 94 & 3.028/3.33$\times 10^{-2}$\\

0.375--0.500 & $0.39^{+0.09}_{-0.08}$  & $5.7^{+2.6}_{-0.4}$  & $9.6^{+1.5}_{-1.3}$
& $4.5^{+1.5}_{-2.8}$ & $34.2^{+1.4}_{-5.7}$ & $4.9^{+8.0}_{-1.9}$ & $-11.9^{+5.2}_{-90.3}$ &$25^{+23}_{-3}$ & 91.8 / 100 & 16.511/8.65$\times 10^{-9}$\\
0.500--0.625 & $0.33^{+0.08}_{-0.07}$  & $6.6^{+0.5}_{-0.4}$  & $6.8^{+0.9}_{-0.7}$
& $3.4^{+0.9}_{-0.8}$ & $34.8^{+0.6}_{-0.6}$ & $4.4^{+1.2}_{-1.1}$ & $-22.9^{+6.3}_{-9.6}$ & $< 31$ & 131.4 / 126 & 53.190/2.78$\times 10^{-22}$\\
0.625--0.750 & $0.37^{+0.08}_{-0.07}$  & $6.5^{+0.5}_{-0.4}$  & $7.9^{+0.1}_{-0.1}$
& $3.9^{+0.9}_{-0.8}$ & $34.2^{+0.6}_{-0.6}$ & $4.7^{+1.2}_{-1.0}$ & $-25.7^{+6.7}_{-10.2}$ &$39^{+5}_{-16}$ & 85.0 / 105& 71.710/ 2.60$\times 10^{-25}$\\
0.750--0.875 & $0.28^{+0.06}_{-0.06}$  & $7.0^{+0.8}_{-0.6}$  & $6.8^{+0.8}_{-0.6}$
& $2.8^{+1.0}_{-0.8}$ & $34.7^{+0.7}_{-0.7}$ & $5.9^{+1.7}_{-1.2}$ & $-31.5^{+10.2}_{-19.0}$ & $37^{+7}_{-23}$&109.8 / 88& 36.737/1.73$\times 10^{-15}$\\
0.875--1.000 & $0.38^{+0.08}_{-0.08}$  & $7.6^{+2.2}_{-1.3}$  & $7.8^{+0.8}_{-0.7}$
& $1.8^{+1.2}_{-0.7}$ & $32.0^{+3.6}_{-16.4}$ & $10.2^{+8.6}_{-6.0}$ & $-48.5^{+61.5}_{-52.4}$ &$31^{+35}_{-5}$& 89.2/ 86& 9.099/ 2.70$\times 10^{-5}$\\
\enddata
\tablenotetext{a}{Referring to equation (1), and defined at 1 keV in units of photons keV$^{-1}$ cm$^{-2}$ s$^{-1}$.}
\tablenotetext{b}{Normalization of the Gaussian is defined in units of photons cm$^{-2}$ s$^{-1}$.}
\tablenotetext{c}{The energy and the width of the Fe-line are fixed at 6.62 keV and $\sigma_{Fe}=0.37$ keV, respectively.}
\tablenotetext{d}{Referring to equation (4), and PCI is probability of chance improvement of the $\chi^2$ by adding a new component.}
\end{deluxetable*}
\begin{figure*}
\begin{center}
 \resizebox{1.0\columnwidth}{!}
  {\includegraphics[width=17cm,height=25cm]{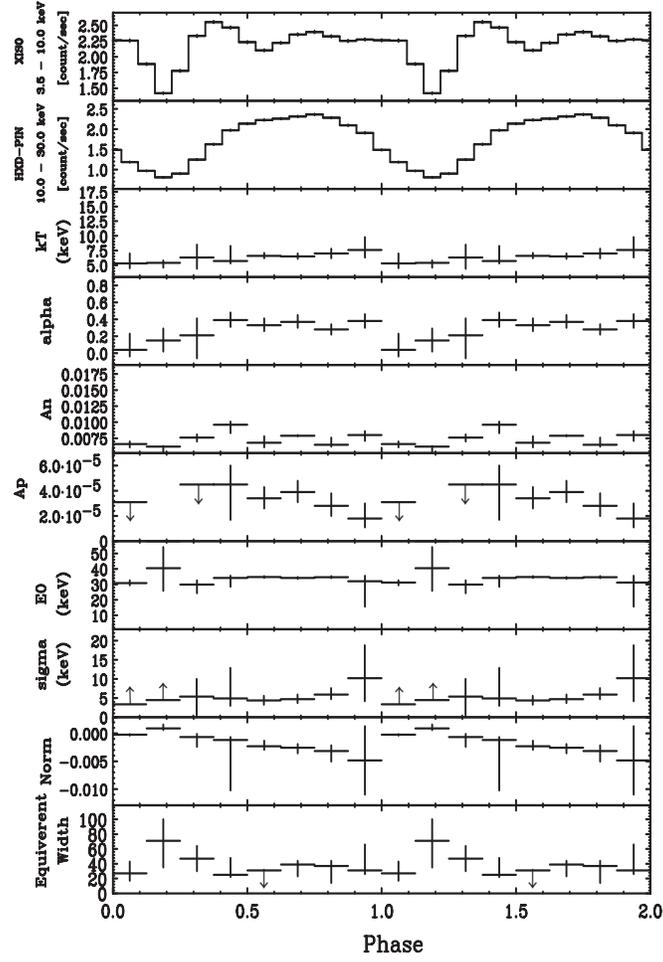}}
 \caption{Fit parameters from as a function of pulse phase for 4U 1626$-$67. Best-fit values by NPEX plus Gaussian model (Table 4) are plotted.}
 \label{fig:f}
\end{center}
\end{figure*}

\begin{figure}
\begin{center}
 \resizebox{1.0\columnwidth}{!}
  {\includegraphics[angle=270]{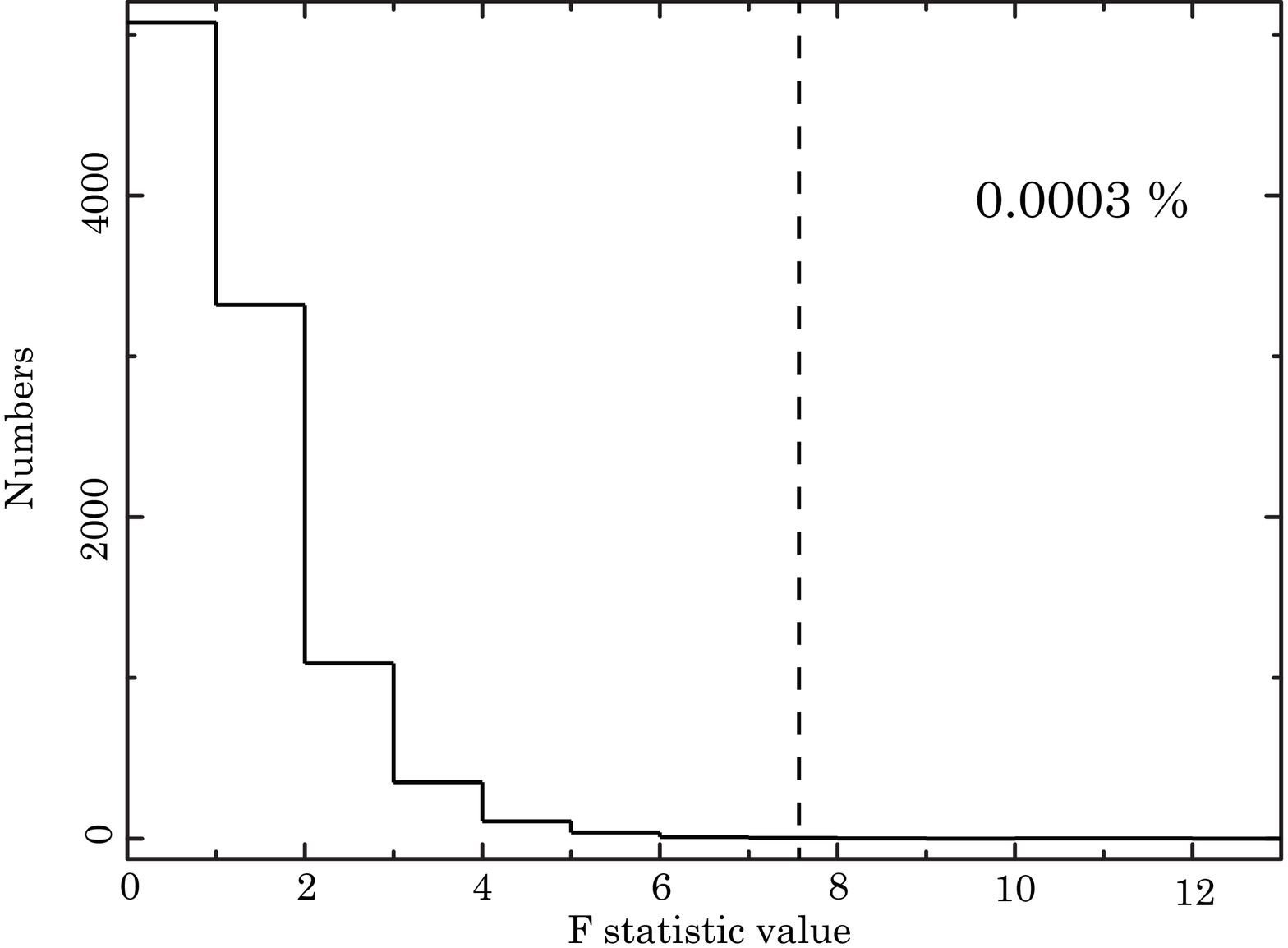}}
 \caption{Observed value of the $F_\chi$ plotted as dashed line against probability histogram of $F_\chi$ obtained by 10,000 simulations (See the text).  }
 \label{fig:}
\end{center}
\end{figure}

Since the GABS model reproduces an absorption feature, we fit the
phase-resolved spectra by the NPEX plus a Gaussian model in which we
allowed the Gaussian normalization to be both positive and negative.
The derived parameters are shown at Table 4 and Figure 10. The results
indicates that the emission line produces the residuals from the NPEX
fitting in Figure 8(a). We can evaluate how much the additional term
improved the fit using the F-test for additional terms
\citep{bev69}. The F statistical value, hereafter called F$_\chi$, is
defined as
\begin{equation}
{F_\chi = \frac{(\chi_{1}^{2} - \chi_{2}^{2})/(\nu_{2} - \nu_{1})}{\chi_{2}^{2}/\nu_{2}}. 
}
\end{equation}
Each suffix denotes model 1 and 2, NPEX only and NPEX + Gaussian
model, respectively. The derived probability from $F_{\chi}$ also
summarized at Table 5. The results show that the data support the
presence of the emission line in the dim phase with $F_\chi$ of
7.688, corresponding to a probability of chance improvement of the
$\chi^2$ of 1.9 $\times$ 10$^{-4}$. The residuals of the fitting are
shown in Figure 8(c). The obtained resonance energy, $E_a =
40.5^{+13.7}_{-14.7}$ keV, is statistically consistent with those
observed in the other phases.

\begin{figure*}
\epsscale{1.1}
\begin{center}
\plottwo{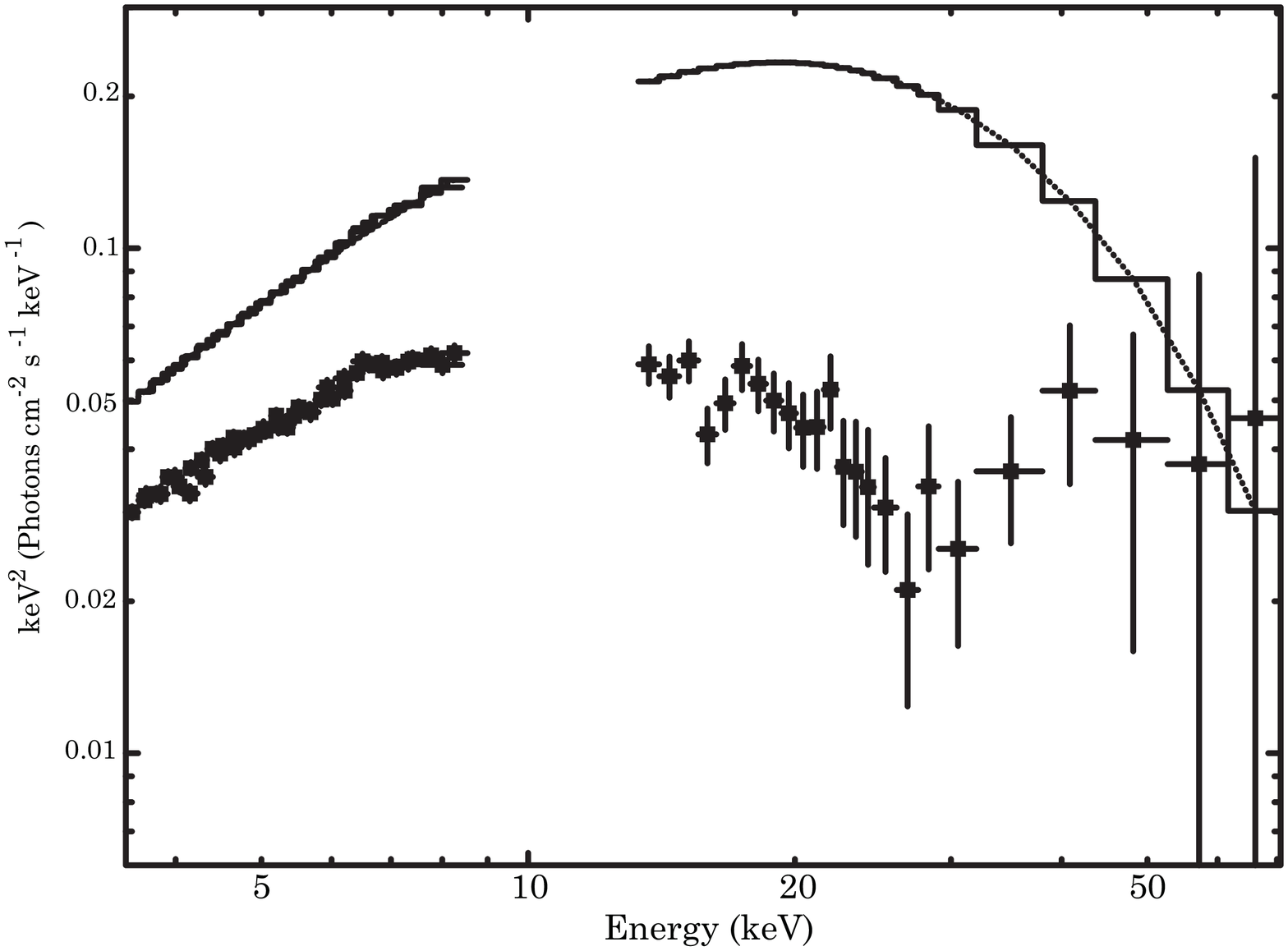}{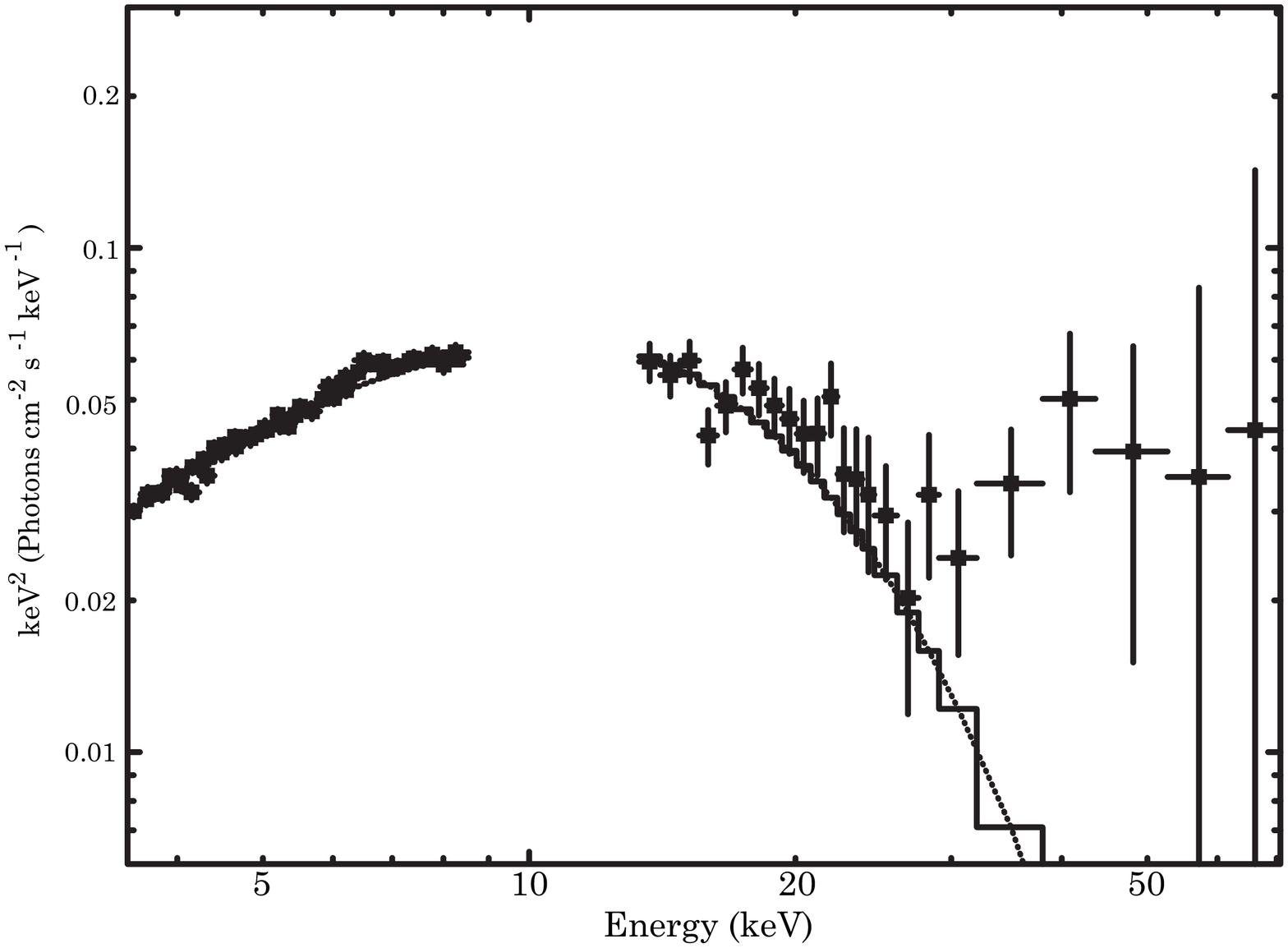}
\caption{The dim phase $\nu F \nu$ spectra of the 4U 1626$-$67 obtained with  XIS and HXD. The NPEX-only ($A_p$ = 0) models are shown in the histogram. (Left) When fitted by NPEX ($A_p$ = 0) with GABS model and (right) when fitted by NPEX ($A_p$ = 0) and Gaussian models.}
\end{center}
\end{figure*}

To examine the chance probability of improvement of the derived
F-value, we simulate the spectra following a procedure similar to
section 5.2 of \citet{pro02}. First, we simulate 10,000 data sets
using "fakeit" in XSPEC Version12 according to the NPEX ($A_p = 0$)
model with parameters fixed at best fit values listed in Table 2, same
instrumental response, background files and same exposure time. Each
data set is binned using same binning as that used to real
data. Second, we fit the simulate data using only NPEX ($A_p = 0$) and
NPEX ($A_p = 0$) + Gaussian model allowing their normalization to be
negative. Third, we calculate the $F_\chi$ derived by these two model
fittings. Last, we evaluated the number of simulated spectra with
$F_\chi$ exceeding the 7.688, which is derived by fitting of real data
to estimate the approximate probability. The derived histogram of the
$F_\chi$ is shown at Figure 11 and the evaluated chance probability is
3.0 $\times$ 10$^{-4}$ which is consistent with the $F_\chi$ value
obtained from real data using equation (4). In order to evaluate the
effect of systematic error of HXD-PIN non-X-ray-background model, we
also refit the dim phase spectrum taking account into the 3 \%
systematic errors of HXD-PIN. The $F_\chi$ is
$7.688^{+2.523}_{-2.490}$ and the probability is between $2.91 \times
10^{-3} $and $1.53 \times 10^{-5}$. Therefore, even though the NXB
level of HXD-PIN is underestimated, we have possibly detected the
emission line at the cyclotron resonance energy in the dim phase.
Although there is a $\sim$ 85 \% possibility that the structure in the
dim phase is produced by absorption feature, we stress that the
continuum requires a hard component with a large cutoff energy, in
addition to a wide and deep absorption feature, as seen in Figure 12.
Furthermore, the $kT$ value at the dim phase matches those at the
other phases with the emission model, whereas a large jump in the
value is required with the absorption model, as shown in Figure 9.

\begin{figure}[!t]
\begin{center}
 \resizebox{1.0\columnwidth}{!}
  {\includegraphics[width=20cm,height=12cm]{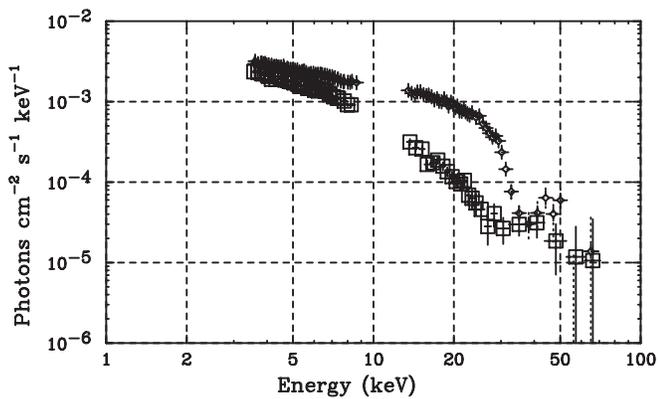}}
 \caption{The unfolded spectrum of 4U 1626$-$67. Crosses denote on-phase spectra ($\phi = 0.625$--0.750). Squares denote off-phase spectra ($\phi = 0.125$--0.250).  }
 \label{fig:g}
\end{center}
\end{figure}

\begin{figure}
\begin{center}
 \resizebox{1.0\columnwidth}{!}
  {\includegraphics[angle=270]{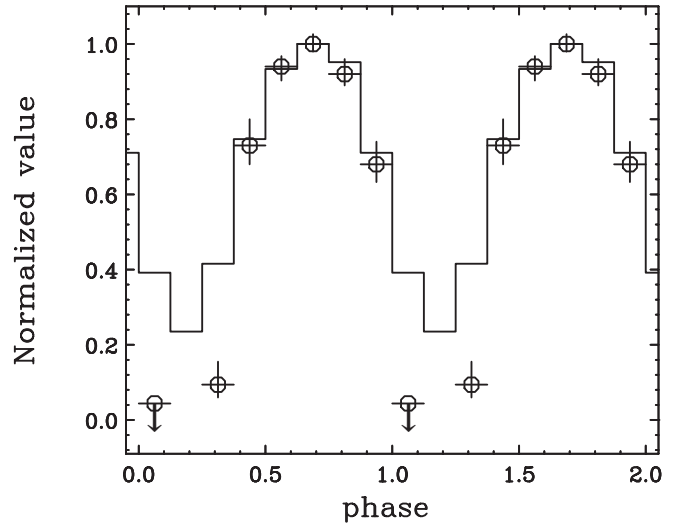}}
 \caption{The plotted line is the background-subtracted pulse profile in the 10--30 keV energy band. The circles denote the best-fit values of the normalization of the positive power-law term (i.e., the Wien hump) when fitting by NPEX model, presented as a function of the pulse phase (see text for details). Values are normalized by each maximum value. }
 \label{fig:h}
\end{center}
\end{figure}

\section{Discussion} 
We observed the 4U 1626$-$67 pulsar with the {\it{Suzaku}} satellite
in March 2006. The pulse period determined by the HXD-PIN of
{\it{Suzaku}} is $P_{(\rm Suzaku)}$ = $ 7.67795(9)$ s, and this value
agrees with that reported by \citet{cam10}. In the pulse
phase-averaged spectra, the fundamental CRSF is clearly detected at
$37.4^{+1.0}_{-0.9}$ keV (GABS model) and the continuum emission is
better reproduced by the NPEX model than the ECUT
model. Phase-resolved spectral analyses consistently show strong phase
dependence, both in the continuum emission and in the CRSF,
particularly in the dim phase. Quantitatively, the positive component
of the NPEX model vanishes near the dim phase, and the CRSF in the dim
phase is reproduced by using a Gaussian shaped emission line model.
The statistical significance of the emission line using the
\citet{bev69} F-test routine is $2.91 \times 10^{-3}$ and $1.53 \times
10^{-5}$ taking account into the systematic errors in the background
evaluation of HXD-PIN. The results means that even if the the real NXB
level of HXD-PIN is 3\% higher, which is the worst case of
underestimating NXB level, the significance level is still around
3$\sigma$ level. Thus, we have possibly detected the emission line at
the cyclotron resonance energy in the dim phase.

In this section, we discuss the nature of the accretion column on the
magnetic pole of the neutron star that produces a cyclotron emission
line at the dim phase rather than an absorption feature.

What is the origin of the emission feature in the dim phase?
\citet{nag81} pointed out that if the photons originate from an
optically thick accretion column (with a Thomson optical depth $\tau_T
> 1$) without temperature and density gradients, the CRSF is expected
to be an absorption feature with Comptonized spectra. In contrast, if
the photons originate from an optically thin accretion column ($\tau_T
< 1$), the CRSF is expected to be an emission feature with a free-free
continuum. To compare this theoretical model with the {\it{Suzaku}}
spectra, the unfolded spectra in the on-phase ($\phi = 0.625$--0.750)
and the dim phase ($\phi = 0.125$--0.250) are shown in Figure
13. Qualitatively, our observed results are consistent with the
theoretical predictions; the Wien hump disappears with the appearance
of an emission-like feature. This resemblance suggests that we
observed the cyclotron resonance scattering emission feature caused by
collisional excitation of a Landau level, followed by radiative
deexcitation. The deciding factor as to whether we observe absorption
spectra or emission spectra is thus the optical depth of the accretion
column, and for the case of 4U 1626$-$67, since change occurs
according to the spin phase, changes in optical depth arise from
changes in viewing angle.

The optical depth, $\tau$, is generally expressed as
\begin{equation}
{\tau = n_{e}\sigma l.
 }
\end{equation}

Here, $n_e$ is the electron number density, $\sigma$ is the scattering
cross section and $l$ is the plasma length-scale through which a photon passes. In a strong magnetic field, where the strength of the field is
greater than the thermal energy, the scattering cross section depends
on the direction of photons relative to the magnetic field
vector. According to \citet{her79}, the scattering cross section of
photons averaged over the energy and the (ordinary and extraordinary)
modes is described by
\begin{equation}
{\sigma_{\parallel} \approx \sigma_{T} \biggl(\frac{\bar{E}}{E_{a}}\biggr)^2,
}
\end{equation}
when the photons have energies far less than $E_a$, and they propagate
parallel to the magnetic field. Here, $\sigma_{T}$ is the Thomson
scattering cross section and $\bar{E}$ is the mean photon
energy. Thus, the optical depth propagating parallel to the magnetic
field, $\tau_{\parallel}$, is
\begin{equation}
{\tau_{\parallel} \approx n_{e} \sigma_{T} \biggl(\frac{\bar{E}}{E_{a}}\biggr)^2 l_{\parallel}.
}
\end{equation}
In contrast, the averaged cross section for photons propagating
perpendicular to the magnetic field, $\sigma_{\perp}$, can be approximated by
\begin{equation}
{\sigma_{\perp} \approx \sigma_{T},
}
\end{equation}
and the optical depth propagating perpendicular to the magnetic field,
$\tau_{\perp}$, is
\begin{equation}
{\tau_{\perp} \approx n_{e} \sigma_{T} l_{\perp}.
}
\end{equation}
Therefore, from equations (7) and (9) we see that the optical depth is
anisotropic, and the probability of Comptonization of photons depends
on the direction that the photons are propagating through the plasma.
This anisotropy of optical depth in the Comptonization process is
expected to cause the phase dependence of the Wien hump.

To verify the phase dependence of the Wien hump, we again fit the
phase-resolved spectra by the NPEX model, except for the dim phase,
where the $kT$ value are instead fixed at the phase-averaged value
(Table 1) by assuming thermal equilibrium. Here, the CRSF is described
by the GABS model and the spectral fits are found to be
acceptable. Figure 14 shows the background-subtracted pulse profile
for the 10--30 keV energy band and compares this with values of
$A_{p}$ plotted as a function of the pulse phase. These two plots
clearly agree. In other words, X-ray fluxes in the 10--30 keV range
are represented by only the Wien hump. Since, as described above,
radiation is caused by the optical depth of scattering, the dim phase
corresponds to the optically thin phase. Therefore, this result is
explained correctly by our prediction.

To estimate the value of $\tau_{\parallel}$ qualitatively, we use the
simple assumptions that the plasma is at thermal equilibrium and that
the accretion column is homogeneous. Under these conditions, $\bar{E}$
is equal to the electron temperature. Therefore, the ratio between
$\tau_{\parallel}$ and $\tau_{\perp}$ is estimated by using the
observed $kT \sim 6.71$ keV and the $E_a \sim 37.4$ keV derived from
fitting the phase-averaged spectra by the NPEX with GABS model (Table
1):
\begin{equation}
\frac{\tau_{\parallel}}{\tau_{\perp}} \approx \biggl(\frac{\bar{E}}{E_a}\biggr)^2 \frac{l_{\parallel}}{l_{\perp}} \approx \biggl(\frac{kT}{E_a}\biggr)^2 \frac{l_{\parallel}}{l_{\perp}} = 3.2 \times 10^{-2} \frac{l_{\parallel}}{l_{\perp}}.
\end{equation}
Thus, if the difference between $l_{\parallel}$ and $l_{\perp}$ is
small ($l_{\parallel} \sim l_{\perp}$), the optical depth for the dim
phase spectra is approximately one or two orders of magnitude lower
than that in the bright phase. If we assume that the Thomson optical
depth of the bright phase is around 10, the cyclotron resonance
scattering absorption feature can be seen as an emission feature at a
certain viewing angle.

In the future, we plan to further investigate the radiative transfer
of photons in the accretion column through theoretical studies that
take into account emission power.

The authors appreciate very much the many constructive
comments from the anonymous referee. We thank all of the {\it{Suzaku}}
team members for their dedicated support in the satellite operation
and calibration. W.B.Iwakiri is a Research Fellow of the Japan Society
for the Promotion of Science (JSPS).

\end{document}